# Direct visualization of magnetic correlations in frustrated spinel ZnFe$_2$O$_4$

Jonas Ruby Sandemann[1], Thomas Bjørn Egede Grønbech[1], Kristoffer Andreas Holm Støckler[1], Feng Ye[2], Bryan C. Chakoumakos[2], Bo Brummerstedt Iversen[1,*]

[1]Center for Integrated Materials Research, Department of Chemistry and iNANO, Aarhus University, DK-8000 Aarhus C, Denmark

[2]Neutron Scattering Division, Oak Ridge National Laboratory, Oak Ridge, Tennessee 37831, USA

*Corresponding author: bo@chem.au.dk

## Abstract

Magnetic materials with the spinel structure (A$^{2+}$B$^{3+}_2$O$_4$) form the core of numerous magnetic devices, but ZnFe$_2$O$_4$ constitutes a peculiar example where the nature of the magnetism is still unresolved. Susceptibility measurements revealed a cusp around $T_c = 13$ K resembling an antiferromagnetic transition, despite the positive Curie-Weiss temperature determined to be $\Theta_{CW} = 102.8(1)$ K. Bifurcation of field-cooled and zero-field-cooled data below $T_c$ in conjunction with a frequency dependence of the peak position and a non-zero imaginary component below $T_c$ shows it is in fact associated with a spin-glass transition. Highly structured magnetic diffuse neutron scattering from single crystals develops between 50 K and 25 K revealing the presence of magnetic disorder which is correlated in nature. Here, the 3D-mΔPDF method is used to visualize the local magnetic ordering preferences, and ferromagnetic nearest-neighbor and antiferromagnetic third nearest-neighbor correlations are shown to be dominant. Their temperature dependence is extraordinary with some flipping in sign, and a strongly varying correlation length. The correlations can be explained by orbital interaction mechanisms for the magnetic pathways, and a preferred spin cluster. Our study demonstrates the power of the 3D-mΔPDF method in visualizing complex quantum phenomena thereby providing a way to obtain an atomic scale understanding of magnetic frustration.

# I. INTRODUCTION

Ferrites that adopt the spinel structure ($Fd\bar{3}m$, AFe$_2$O$_4$) constitute an important class of magnetic materials [1] which combine useful magnetic and electronic properties with a high degree of tailor ability through compositional control, particle size, and defects. This versatility has allowed spinel ferrites to also find applications within many other fields [2] such as microwave devices [3], catalysis [4], gas sensing [5], water purification [6] and pharmacy [7].

ZnFe$_2$O$_4$ has been widely studied but considerable disagreement remains about the magnetic nature of the system [8,9]. Pristine ZnFe$_2$O$_4$ adopts a defect-free normal spinel structure where oxygen forms a distorted cubic closest-packing with Zn$^{2+}$ cations placed at the 8$a$ tetrahedral (A) site and Fe$^{3+}$ at the 16$d$ octahedral (B) site [10]. The unit cell can be constructed from alternating stacking of two building blocks, as shown in Fig. 1(a)-(c). Since Zn$^{2+}$ has no magnetic moment the magnetic ions solely reside on the B sites, which constitutes a pyrochlore lattice of corner-sharing tetrahedra (Fig. 1(d)) conducive to magnetic frustration [11-13].

Early investigations based on specific heat measurements [14] and neutron powder diffraction [15] suggested that ZnFe$_2$O$_4$ is paramagnetic at room temperature with an antiferromagnetic transition occurring around $T_N = 10$ K. However, short-range order has been observed much above the proposed Néel temperature and found to coexist with the long-range order below $T_N$ [16], and Usa *et al.* found no sign of long-range order down to 1.5 K in pristine ZnFe$_2$O$_4$ indicating an intrinsically frustrated magnetic system [17]. This was further corroborated by the presence of structured neutron diffuse scattering at low temperatures [18]. Theoretical analysis of these diffuse scattering data predicts ferromagnetic nearest-neighbor interaction ($J_1$) and antiferromagnetic third-neighbor interaction ($J_3$) [19]. A ferromagnetic nearest-neighbor coupling is in stark contrast to the earlier descriptions of ZnFe$_2$O$_4$ as an antiferromagnet, but it corroborates the Curie-Weiss temperature of 120 K [18] and excess small angle neutron scattering observed at 89 K [20]. Whether the system constitutes a spin-glass phase below $T_N$ is also debated with susceptibility measurements giving evidence both for [21] and against [18] this interpretation. Despite all the evidence for magnetic frustration, a recent report still claims ZnFe$_2$O$_4$ to be an antiferromagnet [9].

ZnFe$_2$O$_4$ clearly constitutes a delicate magnetic system, where competing orbital interactions between atoms locked in a specific geometrical configuration are easily perturbed. Here we probe the elusive magnetism in ZnFe$_2$O$_4$ using the recently developed three-dimensional magnetic difference pair distribution function (3D-mΔPDF) analysis, which is a model free method for reconstructing and visualizing magnetic correlations [24,25]. Newly synthesized high-quality single crystals (low cation inversion and no interstitials) were used in the present study (see Supplemental Material).

## II. RESULTS

### A. Spin glass behavior

Magnetization measurements were performed to investigate the macroscopic magnetic behavior of the $ZnFe_2O_4$ single crystals. The static magnetic susceptibility χ was determined using field-cooled (FC) and zero-field-cooled (ZFC) magnetization (Fig. 2(a)). The ZFC data resembles the expected behavior of an antiferromagnet with a Néel temperature defined by the cusp in the susceptibility at 13.1 K (Fig. 2(a) inset). However, modelling of the temperature dependence of the inverse susceptibility between 260-400 K gives a Curie-Weiss temperature of $\Theta_{CW} = 102.8(1)$ K (Fig 2a) indicating that ferromagnetic correlations dominate the spin-spin interactions at room temperature. The large discrepancy between $\Theta_{CW}$ and the apparent transition temperature, $T_c = 13.1$ K, hints at a great degree of frustration, as quantified in the high frustration index $f = |\Theta_{CW}|/T_c = 7.86(6)$ [26]. Furthermore, the bifurcation of the FC and ZFC susceptibility observed below $T_c$ is one of the signatures of a spin-glass transition [27,28]. The effective moment of the $Fe^{3+}$ ions was calculated from the slope of the Curie-Weiss fit to be $\mu_{eff} = 3.04(5)$ BM, which is only half the spin-only value for high spin $Fe^{3+}$ (5.9 BM) indicating covalency in the chemical bonding.

AC susceptibility measurements were conducted in a frequency range spanning close to four orders of magnitude to further investigate the potential spin-glass transition. The results are displayed in Fig. 2(b) showing a clear shift in the cusp of the real component of the AC susceptibility (χ′) as the frequency increases, another characteristic of a spin-glass transition. The peak moves from 13.1 K at 46 Hz to just below 14 K at 9984 Hz (Fig. 2(b) inset), but this frequency dependence is only seen below $T_c$. The peak in χ′ is likewise associated with a sharp maximum in the imaginary component of the AC susceptibility (χ′′), which is otherwise zero above $T_c$. Non-zero values of χ′′ are related to dissipative processes, such as irreversibility and relaxation processes in the spin-glass phase [27,28]. Heat capacity measurements confirm the lack of a magnetic phase transition at 13 K (see Supplemental Material).

In summary, the physical properties strongly suggest that $ZnFe_2O_4$ is a spin-glass and that the cusp in the susceptibility at 13 K is associated with spin-freezing rather than a paramagnetic to antiferromagnetic transition.

### B. Local magnetic correlations

Single crystal neutron scattering was used to gain insights into the apparent magnetic disorder in $ZnFe_2O_4$. Neutron diffraction patterns measured at various temperatures between 50 K and 1.5 K are shown in Fig. 3(a)-(e). Diffuse scattering develops at lower temperatures with highly structured lines appearing between 50 K and 25 K signifying that the disorder is locally correlated and non-random.

Recently the three-dimensional magnetic difference pair distribution function (3D-mΔPDF) method was developed, and used to directly reconstruct the magnetic correlations in the frustrated spin-glass mineral bixbyite [24,25]. The 3D-mΔPDF is the inverse Fourier transform of the unpolarized magnetic diffuse neutron scattering cross-section and is a direct space function with peaks corresponding to interatomic vectors separating more or less magnetization density pointing in the same direction compared to the average periodic structure [25]. When there is no long-range magnetic order, as is the case for $ZnFe_2O_4$, the interpretation becomes comparatively simple; positive (negative) peaks correspond to vectors separating magnetization density pointing in the same (opposite) direction.

Fig. 3(f) and g show two planes of the isolated diffuse scattering obtained from the 1.5 K dataset along with the *a-b* plane through the origin of the Fourier transformed 3D-mΔPDF. The well-defined peaks immediately show that magnetic correlations exist and the lack of significant features at longer distances highlights that the correlations are highly localized.

The local ordering can be elucidated from the sign and position of the peaks in the 3D-mΔPDF combined with knowledge of the atomic structure. Fig. 3(j)-(l) shows slices in the *a-b* plane of the 3D-mΔPDF of the 1.5 K dataset at three different positions along the *c*-axis, and in Fig. 3(i) the spinel unit cell is shown with only the octahedral sites. The interpretation of the 3D-mΔPDF follows from defining one of the sites as the origin (0). Then red and blue atoms tend to have co-parallel and anti-parallel spins compared to 0 respectively. The nearest neighbours (nn1) are all separated by a vector (0, ¼, ¼) and its equivalents ($a \approx 8.43$ Å). For z=0 Å and z=2.11 Å these are clearly seen as positive peaks indicating a preference for co-alignment of neighbouring spins. The separation vector to the second (nn2) and third (nn3) nearest neighbors are (¼, ¼, ½) and (0, ½, ½) and their equivalents, respectively, and both show up as negative peaks in the 3D-mΔPDF revealing a preference for antiparallel alignment with nn2 and nn3.

From this analysis, the frustrated nature of the magnetism is apparent; it is not possible to simultaneously have ferromagnetic alignment between nearest neighbors and antiferromagnetic alignment with next-nearest neighbors, only complete ferromagnetic ordering is compatible with co-alignment of all nearest neighbors. In addition, the antiferromagnetic interaction of third nearest neighbors by itself is frustrated since these sites form a face-centered cubic sublattice, an edge-sharing tetrahedral network, which is incompatible with long range antiferromagnetic order, resulting in multiple levels of frustration being at work in this material. The relative tetrahedral arrangement of the octahedral nn3 atoms is illustrated in Fig. 1(d).

## C. Correlations across the spin-glass transition

A quantitative analysis can give further insights into the spatial and thermal dependence of the correlations. The *a-b* plane through the origin of the 3D-mΔPDFs is plotted from 1.5 K to 50 K in Fig. 4(a)-(e). The increase in correlation length and strength with decreasing temperature is evident, with only weak ferromagnetic nn1 coupling at 50 K evolving into a more complicated picture described above at 1.5 K. The correlation length along any direction can be evaluated by the extent to which significant features exist in the 3D-mΔPDF. The peak intensities for the 20 shortest Fe-Fe distances were integrated (Fig. 4(f)) from which the correlation length can be estimated at around 15 Å at 1.5 K, and from the 3D-mΔPDF itself this appears to be fairly isotropic. Above 17 K there are no strong correlations beyond nn3 at 5.9 Å, but a few weak correlations extending to nn8 at 10.3 Å. This reveals a strong temperature dependence of the correlation length in the vicinity of $T_c$. Interestingly, there are correlations that change sign with increasing temperature (marked by asterisks in Fig. 4(f)). A switch from a ferro- to an antiferromagnetic interaction with changing temperature may suggest competing orbital interactions.

To get a better understanding of the temperature behavior of the individual correlations the normalized intensity of five selected peaks is plotted as a function of temperature in Fig. 4(g). The rate of decay with temperature varies significantly with the longer correlations generally dying out faster than shorter ones. Comparing for example the 14.6 Å (nn14) peak to the 3.0 Å peak (nn1), the former has dropped more than 90 % at 17 K whereas the latter retains more than 70 % of its 1.5 K value at this temperature. Thus, it is not merely the strength of the correlations that diminishes with temperature, the correlation length decreases as well. This is in contrast to the findings for bixbyite, which showed a nearly constant correlation length with temperature [24]. The varying behavior of the different peaks with temperature corroborates a complex set of competing magnetic interactions in $ZnFe_2O_4$.

The integrated intensities reveal that the nn3 correlation is stronger than the nn1 correlation at low temperature: |nn3|/nn1 = 1.1 at 1.5 K (Fig. 4(h)). The nn3 distance is twice that of the nn1 making this a rather unintuitive result, but it is in line with earlier findings [19]. The top panel of Fig. 4(h) shows the dramatic temperature dependence of the ratio between |nn3| and nn1 with a steep, relatively linear decline from 1.5 K to 50 K and then a minor drop up to 150 K. The ratio of 1.1 at 1.5 K is significantly lower than the similar ratio $J_3/|J_1| = 4.0$ at 1.5 K determined theoretically by Yamada *et al*. [19]. However, the integrated correlation peak intensities are not the same as the *J*-couplings and as such cannot be expected to be identical. The decrease in |nn3|/nn1 with temperature is also much more gradual where there is still significant nn3 contribution at 50 K, compared to Fig. 7 in Yamada *et al.* where $J_3/|J_1|$ is close to 0 already at 30 K. The bottom panel of Fig. 4(h) shows the striking similarity in temperature dependence of nn1+nn3 and nn2 alluding to similar underlying mechanisms.

## III. DISCUSSION

The magnetic interactions are frustrated at all temperatures, as evidenced by the diffuse scattering occurring significantly above $T_c$. The increase in |nn3|/nn1 with decreasing temperature is, in addition to being emblematic of the general increase in correlation length, an indication of a change in the nature of the local spin clusters with temperature. At higher temperatures they are dominated by the shortest-range nn1 interactions which presumably manifests as small ferromagnetic clusters that behave basically independent of each other, as seen from the 3D-mΔPDF at 50 K. As the temperature decreases the relative strength of the antiferromagnetic nn3 correlation increases causing the ferromagnetic spin clusters to preferably align oppositely to neighboring clusters. As nn3 becomes comparable to nn1 the clusters resemble ferromagnetic cores that couple antiferromagnetically to a larger shell. The concurrent increase in correlation strength and length may create an energy barrier that is sufficiently large compared to the thermal energy $k_BT$, and this freezes the spins resulting in the observed spin-glass transition. The observed magnetic correlations may be understood from the quantum mechanical exchange interactions in the structure. From the 3D-mΔPDF intensities, the first- and third-nearest neighbor interactions are the most prominent, in agreement with previous theoretical analysis of single crystal neutron diffuse scattering [19]. The exchange mechanisms in transition metal oxides, such as $ZnFe_2O_4$, are often rationalized using the Goodenough-Kanamori-Anderson rules [29-31]. The different exchange pathways and mechanisms following these rules are discussed below and illustrated in Fig. 5.

For nn1 there are two potential pathways: Fe-Fe as direct exchange $J_1^d$ and Fe-O-Fe as superexchange through oxygen at a 90 ° angle $J_1^s$. The half-filled *d*-shell of $Fe^{3+}$ (electron configuration [Ar]$3d^5$) results in a weak antiferromagnetic $J_1^d$ due to the large overlapping distance (Fig. 5(a)), while there are potentially competing ferro- and antiferromagnetic contributions in $J_1^s$. The ferromagnetic term could arise from interaction between the Fe 3*d* $e_g$ and O 2*p* σ orbitals where the orthogonal *p*-orbitals interacting with each $Fe^{3+}$ result in a ferromagnetic coupling between the cations (Fig. 5(b)). An example of a potential antiferromagnetic mechanism is coupling of a 3*d* $e_g$ with a 3*d* $t_{2g}$ orbital through the same O 2*p* σ (Fig. 5(c)) [30,32]. The dominating term is, at least partly, dictated by the lattice parameter, where a relative increase of the ferromagnetic term with increasing lattice parameter has been observed in chromite spinels [33]. From the sign of the nn1 peak it is clear that the total interaction $J_1$ is ferromagnetic revealing that the ferromagnetic $J_1^s$ is the dominant coupling, resulting in preferential co-alignment of nearest neighbor spins.

The nn2 and nn3 pathways appear identical based on the individual bonds traversed, however, their difference lies in the connections they form to neighboring tetrahedra. Both nn2 and nn3 can be connected via two 90° Fe-O-Fe bonds, i.e. the sum of two consecutive $J_1$ interactions. Another possible pathway is through an intermediary tetrahedral Zn, Fe-O-Zn-O-Fe, which also connects to nn2 and nn3, $J_2$ and $J_3$, respectively. A potential exchange mechanism for these pathways is shown in Fig. 5(d). The Fe 3*d* $e_g$ interacts with the O 2*p* σ and an orthogonal O 2*p*-orbital interacts with the empty 4*s* orbital on $Zn^{2+}$. Due to the potential exchange between the O 2*p*-orbitals, the spin that is virtually transmitted to Zn will have the opposite spin of Fe. The same exchange happens from the other Fe to Zn, however, the electrons virtually transferred from O to Zn must have opposite spins due to the interaction being through the empty 4*s* orbital of Zn. The net result of these interactions is that the two Fe must have an antiferromagnetic coupling. This agrees with the sign of the 3D-mΔPDF peaks for nn2 and nn3 being negative, such that the overall correlation is antiferromagnetic. However, the strength of the correlations differs significantly with the second nearest-neighbor peak being roughly ¼ of the third nearest-neighbor peak at 1.5 K (Fig. 4(f)).

The $J_2$ and $J_3$ pathways only differ in the outgoing bond they utilize after the tetrahedral Zn (Fig. 5(f)). Accordingly, the mechanism in Fig. 5(d) is valid for both, and it is thus expected to contribute equally to the nn2 and nn3 correlations. The discrepancy between the observed correlation strength then must lie in other competing interactions, the origin of which is apparent from Fig 5f. $J_2$ connects Fe atoms in adjoining tetrahedra whereas $J_3$ connects atoms in tetrahedra that are not directly connected. The atoms connected by $J_2$ will consequently share a nearest neighbor which will be ferromagnetically coupled to each other creating a competing interaction for nn2. The similar behavior of nn2 and nn1+nn3 (Fig. 4(h) bottom) reinforces the description of nn2 being governed by a combination of the nn1 and nn3 exchange mechanisms. The downshift in the nn2 curve compared to nn1+nn3 is expected since the competing ferromagnetic coupling would be through a common nearest neighbor, not a direct one, and the competing ferromagnetic correlation would then be weaker than a pure nn1 interaction. Such a competing $J_1$ interaction would also be present in nn3, however, this traverses an additional Fe-O-Fe linkage and would thus be significantly weaker than that in nn2.

Fig. 5(g)-(h) shows two potential preferred spin clusters around any given Fe atom, the former only based on the 3D-mΔPDF while the latter takes the above considerations on exchange into account. Fig. 5(g) shows the surrounding nn1s are co-aligned giving an hourglass shaped ferromagnetic core surrounded by a shell of oppositely aligned nn3s, both those connected through Fe-O-Zn-O-Fe (nn3$_{Zn}$) and through Fe-O-Fe-O-Fe (nn3$_{Fe}$) linkages. The oppositely aligned nn2s are disregarded due to their relatively weak correlation strength. In Fig. 5(h) nn3$_{Fe}$ is omitted as these are more likely to be weakly ferromagnetically coupled based on the exchange pathway (a sum of two $J_1^S$ interactions), and as such are not expected to contribute significantly to the nn3 peaks in the 3D-mΔPDF compared to those connected through Zn. However, since the nn3$_{Zn}$ and nn3$_{Fe}$ vectors are identical they cannot be distinguished in the 3D-mΔPDF. This magnetic structure is incompatible with the spinel lattice, hence the observed frustration and lack of long-range order. This model stands in stark contrast to the dodecamer spin-molecule proposed by Tomiyasu *et al.* [23], which has co-alignment of nn3$_{Zn}$ and opposite alignment of nn3$_{Fe}$.

## IV. Conclusion

The magnetic disorder in ZnFe$_2$O$_4$ has been examined using magnetization measurements along with single crystal neutron diffuse scattering and 3D-mΔPDF analysis. The static susceptibility showed a cusp at $T_c = 13$ K below which FC-ZFC bifurcation was observed, indicating a transition to a spin-glass phase. A positive Curie-Weiss temperature $\Theta_{CW} = 102.8(1)$ K signified dominant ferromagnetic interactions and considerable magnetic frustration ($\frac{\Theta_{CW}}{T_C} \gg 1$). AC susceptibility revealed a clear frequency dependence of the cusp position in χ′. This, together with a lack of evidence of a magnetic phase transition in the heat capacity data, establishes ZnFe$_2$O$_4$ as a spin-glass material.

Single crystal neutron measurements revealed highly structured diffuse scattering at low temperatures. 3D-mΔPDF analysis of the diffuse scattering allowed determination of correlation strengths and signs for the different Fe-Fe nearest neighbors. The magnetic structure is dominated by nearest neighbor ferromagnetic correlations and third nearest neighbor antiferromagnetic correlations. The correlations are very local, with the correlation length peaking at approximately 15 Å at 1.5 K and decreasing significantly as the temperature is increased. The temperature dependence of individual correlations, as well as the relative magnitudes of the first, second, and third nearest neighbor correlations, suggest competing interactions. The antiferromagnetic nature of the second and third nearest neighbor interactions can be explained from a super-exchange mechanism through the empty 4$s$ orbital of Zn. The ratio |nn3|/nn1 increases rapidly with decreasing temperature hinting at a possible mechanism for the spin-freezing as a pinning of the spin-clusters when the nn3 correlations become dominant. Analysis of the detailed 3D single crystal neutron diffuse scattering offers a means to build microscopic understanding of magnetic exchange mechanisms in magnetically disordered materials which is not possible from the Bragg scattering alone.

## Acknowledgements

We gratefully acknowledge prof. Charles Lesher for measuring the crystal stoichiometry. This research used resources at the Spallation Neutron Source, a DOE Office of Science User Facility operated by the Oak Ridge National Laboratory. The study was supported by the Villum Foundation.

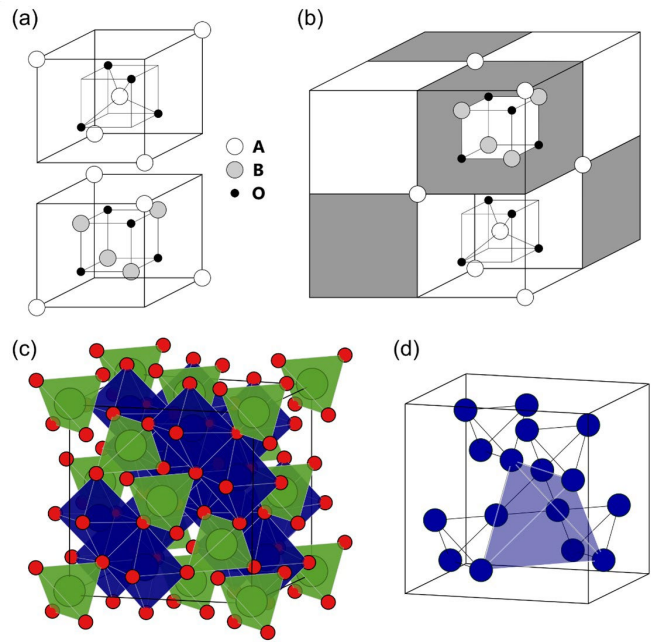

FIG. 1. The spinel structure. (a) The two building blocks of the unit cell with the tetrahedral A block (top) and the octahedral B block (bottom). (b) The unit cell built from alternate stacking of the A and B blocks. (c) Polyhedral model of the spinel unit cell showing the tetrahedra (green) and octahedra (blue) formed by the distorted cubic closest-packed oxygen (red). (d) The corner-sharing tetrahedral network (pyrochlore lattice) formed by the octahedral sites with the tetrahedral coordination of third nearest octahedral neighbors highlighted.

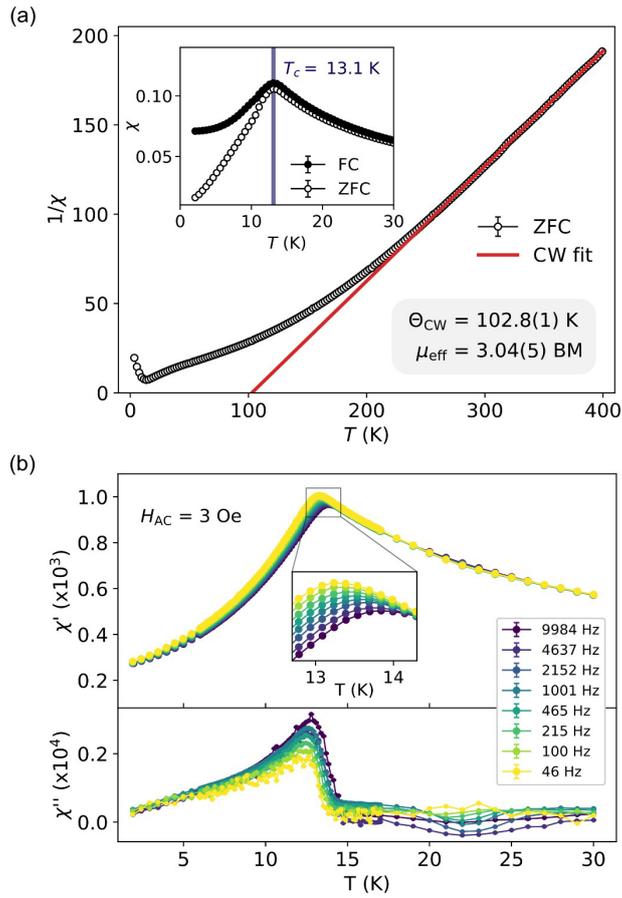

FIG. 2. Magnetization and heat capacity measurements. (a) The inverse static susceptibility as a function of temperature with a Curie-Weiss fit to the linear, high temperature region (T<260 K), along with the resulting Curie temperature and effective magnetic moment. The inset shows the susceptibility as a function of temperature in the low temperature region with a clear cusp around 13 K and bifurcation of the field-cooled (black circles) and zero-field-cooled (white circles) data below this cusp. (b) The real (top) and imaginary (bottom) component of the AC susceptibility as a function of temperature measured in the frequency interval 46-9984 Hz. The variation in the position of the cusp with frequency is highlighted in the inset.

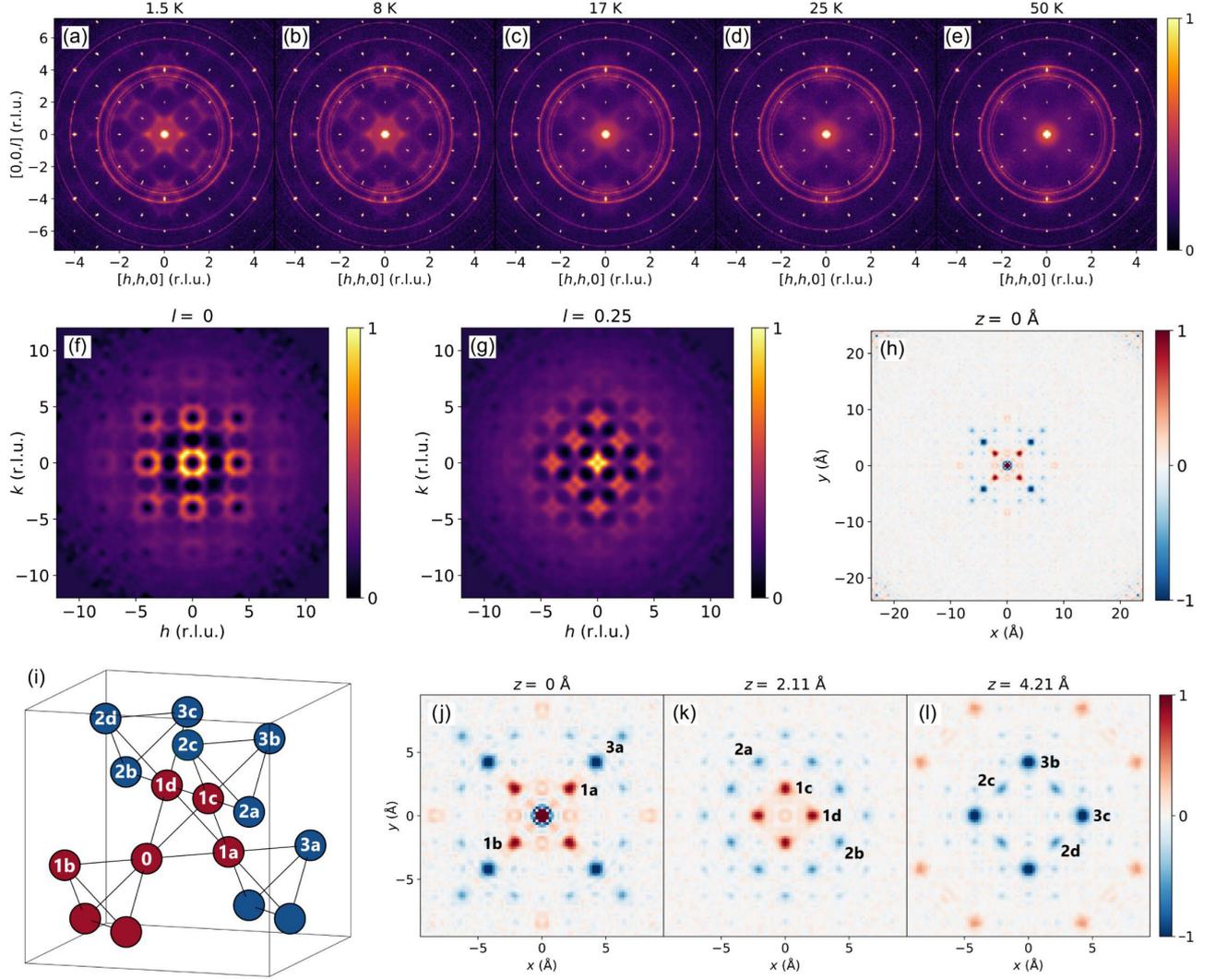

FIG. 3. Single crystal neutron scattering and 3d-mΔPDF analysis. (a)-(e), The (hhk) plane reconstructed directly from the raw frames between 1.5 K and 50 K showing strong condensation of the diffuse scattering into a structured pattern as the temperature is lowered. The powder rings are a result of scattering from the sample holder. (f)-(g), Isolated diffuse scattering signal in the hk plane at l=0 (f) and l=0.25 (g). (h) The a-b plane of the 3D-mΔPDF at z=0 Å showing the short-ranged nature of the local correlations ($a \approx 8.43$ Å). (i) The octahedral sites in the spinel unit cell colored according to the sign of the peak of the corresponding interatomic vector (red for a positive correlation and blue for a negative) with respect to the origin atom (0). The number of the labels correspond to the coordination shell relative to atom 0. (j)-(l), A close look at the correlations in the a-b plane at three different positions along the c axis. The peaks are labelled according to the corresponding atom in (i).

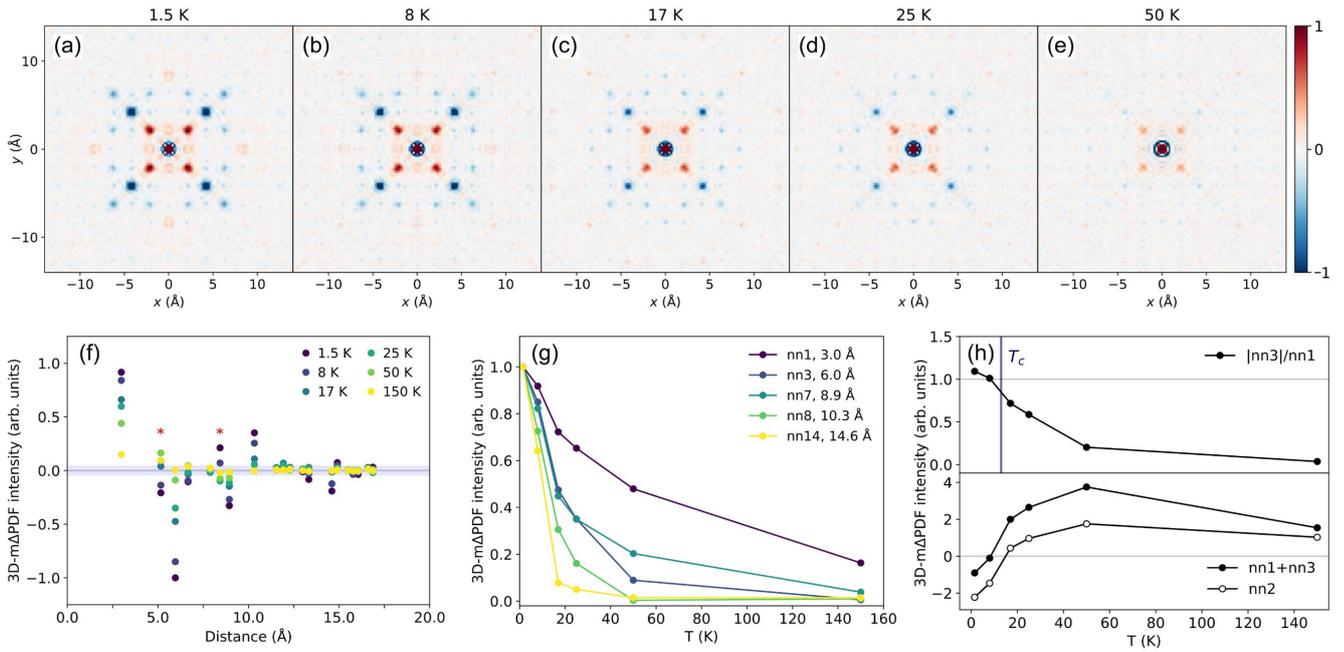

FIG. 4. Quantification of magnetic correlations with temperature. (a)-(e), The *a-b* plane of the 3D-mΔPDF at z=0 Å from 1.5-50 K showing the growth in correlation strength and length as the temperature is lowered. (f) The 3D-mΔPDF integrated peak intensity of the 20 nearest octahedral neighbors as a function of distance for different temperatures. The red asterisks mark significant peaks whose sign of the correlation depends on the temperature. Some peaks switch sign but only within the light blue region where the change in sign is not sufficiently strong to be considered significant based on the level of noise observed in the 3D-mΔPDF. (g) The intensity of five significant peaks as a function of temperature normalized to their 1.5 K value highlighting the varying temperature dependence of the correlations. (h) The top panel shows intensity ratio of the third and first nearest-neighbor peaks as a function of temperature. The critical temperature, as determined from magnetization measurements, is shown by the blue vertical line. The bottom panel compares the temperature behavior of the sum of the first and third nearest neighbor correlations with that of the second nearest neighbor.

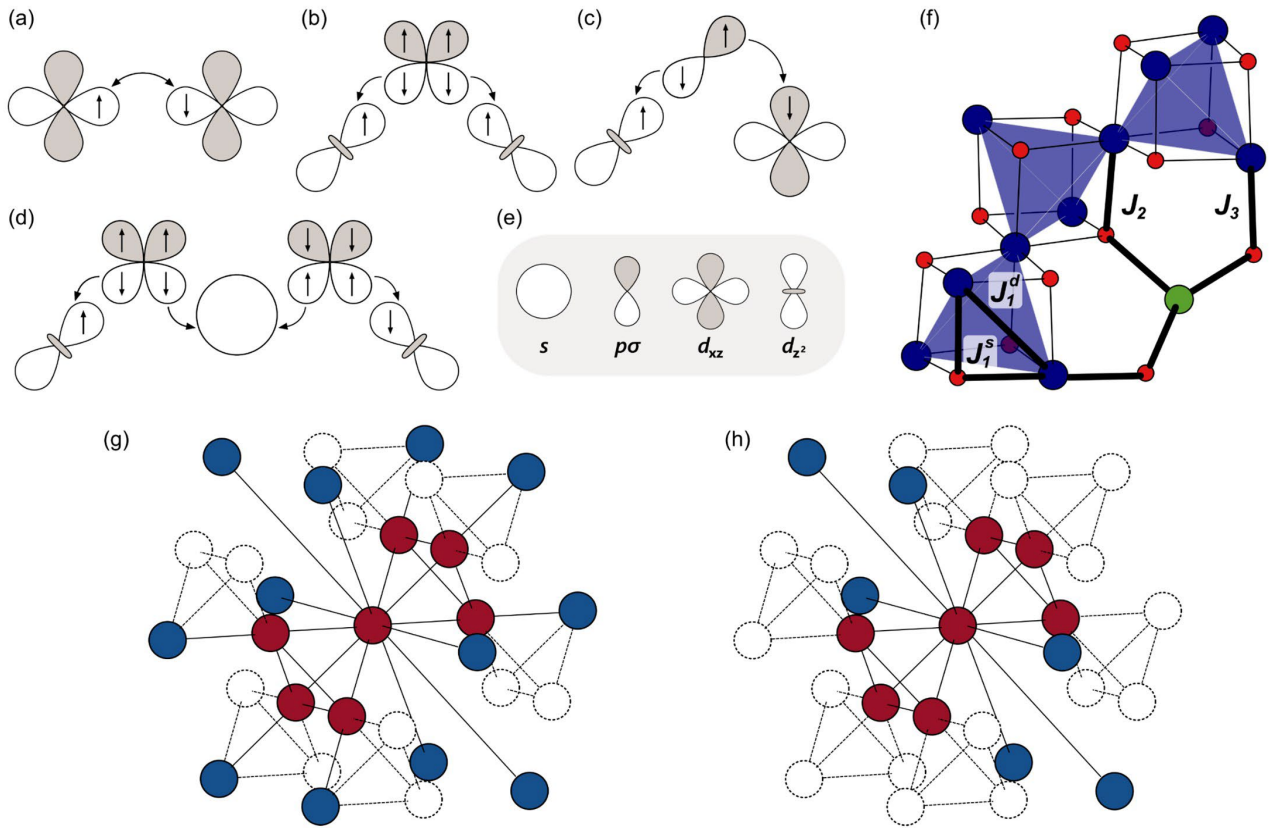

FIG. 5. Exchange mechanisms, pathways, and preferred spin clusters. (a) Antiferromagnetic direct exchange between neighboring Fe 3$d$ t$_{2g}$ orbitals. (b) Ferromagnetic superexchange between neighboring Fe 3$d$ e$_g$ orbitals via orthogonal O 2$p$ σ orbitals. (c) Antiferromagnetic superexchange between neighboring Fe 3$d$ e$_g$ and 3$d$ t$_{2g}$ through the same O 2$p$ σ orbital. (d) Antiferromagnetic superexchange of third nearest neighbor Fe 3$d$ e$_g$ through O 2$p$ σ and an empty Zn 4$s$. (e) Legend of the different types of orbitals illustrated in (a)-(d). (f) A small section of the spinel structure where the different exchange paths have been highlighted. Blue atoms are Fe, red atoms O, and the green atom Zn. (g) The preferred spin cluster around any given Fe atom based solely on the 3D-mΔPDF; all nearest neighbor (red circles) spins co-align and all third nearest neighbor (blue circles) spins with opposite alignment. All nn1, nn2 and nn3 atoms are included but the nn2 atoms (dashed circles) are omitted from the spin cluster description due to the relative weakness of their correlation. (h) The preferred spin cluster where the proposed exchange mechanisms are considered in addition to the 3D-mΔPDF resulting in the third nearest neighbors connected through a nearest-neighbor Fe being omitted in addition to nn2.

# Supporting information

## S1. Methods

### S1.1. Single crystal growth

$ZnFe_2O_4$ single crystals were grown using the flux method. The growth procedure was based on work done by Manzel [1] who mapped out the part of the solubility curve of Zn and Ni ferrite in PbO. Accordingly, PbO(99.99 %, metals basis, Alfa Aesar) was used as the flux material and was mixed with equimolar amounts of $Fe_2O_3$(99.99 %, chemPUR) and ZnO(99.99 %, metals basis, Alfa Aesar) in a Pt crucible using 50 g of PbO and a 4:1 weight ratio of the flux and the spinel components. The crucible was covered with a loose-fitting lid and partially submerged in $Al_2O_3$ sand in a large alumina crucible. The alumina crucible and sand were used as additional thermal mass to mitigate thermal oscillations during the slow cooling to prevent a varying growth rate which would be deleterious to the crystal quality. The Alumina crucible was placed in a Nabertherm L5/14 muffle furnace and subjected to the following temperature program: heating to 1473 K at 600 K/h, then cooling to 1423 K at 10 K/h followed by 1 K/h cooling down to 1173 K where it was held for 24 h after which the furnace was turned off to cool naturally to room temperature. Once the sample had cooled the PbO was leeched with hot $HNO_3$ revealing large, octahedral crystals.

### S1.2. Physical property measurements

Heat capacity and magnetization measurements were conducted using a Quantum Design Physical Property Measurement System (PPMS). For the determination of the heat capacity, an addenda measurement was carried out sampling 50 temperatures distributed logarithmically from 1.7 K to 300 K to determine the heat capacity of the sample holder. Subsequently a 16.931(1) mg $ZnFe_2O_4$ single crystal was added, and the heat capacity measured at 60 points distributed logarithmically from 1.7 K to 300 K. The value at each temperature was taken as the average of three measurements, both for the addenda and sample measurements.

For the magnetization measurements the PPMS was equipped with an Alternating Current Measurement System (ACMS II) susceptometer. Direct current (DC) susceptibility was measured between 1.7 K and 400 K at a field strength of 50 Oe. Data was collected continuously upon heating with each data point being a five second average. From 1.7-10 K the heating rate was 60 K/h and above 10 K it was 120 K/h. The cooling rates were the same in the respective temperature regions. AC susceptibility was measured with a driving field strength of 3 Oe using 10 different frequencies distributed logarithmically between 10 Hz and 10000 Hz (the two lowest frequencies have been omitted from Fig. 2(b) due to excessive noise). The susceptibility was recorded between 1.7 K and 30 K in 1 K increments, except for the region 6-17 K where 74 equidistant data points were collected to get adequate resolution around the cusp.

### S1.3. Neutron Scattering

Single crystal neutron scattering data was collected using two different instruments at the Spallation Neutron Source at Oak Ridge National Laboratory. CORELLI was used to collect elastic diffuse scattering [2,3]. A $ZnFe_2O_4$ single crystal was glued to an aluminum pin and secured further with a thin aluminum foil. A neutron absorbing Cd foil was wrapped around the pin to reduce scattering from the sample holder. The holder was mounted in an orange cryostat and data sets were collected at 1.5 K, 8 K, 17 K, 25 K, 50 K, 150 K and 270 K.

TOPAZ was used to collect diffraction data at 90 K for structural refinements in order to determine the degree of inversion [4].

## S2 Neutron data treatment

### S2.1. Neutron scattering data reduction

To generate the 3D-mΔPDF the magnetic diffuse scattering must first be isolated from the scattering data. This general process was outlined in further detail by Roth *et al*. [5].

Using instrument specific scripts, the UB matrix was determined for each goniometer rotation angle and used to reconstruct the elastic scattering in reciprocal space and normalized to vanadium flux. The reciprocal space reconstruction was done on a $401 \times 401 \times 401$ grid with each axis going from -20 Å$^{-1}$ to 20 Å$^{-1}$. The 270 K data is well into the paramagnetic regime and was subtracted from the lower temperature data sets to remove all scattering contributions expect the low temperature magnetic scattering, such as nuclear Bragg scattering, thermal diffuse scattering, background scattering from the sample holder, etc. The Bragg peaks, however, were not perfectly subtracted with this method, there were still significant residual errors. These were removed using a punch and fill method and due to a slight increase in peak size with *Q*, peaks with $\sqrt{h^2+k^2+l^2} > 6.5$ a punch of $3 \times 3 \times 3$ pixels was used whereas for the peaks with $\sqrt{h^2+k^2+l^2} < 6.5$ a slightly smaller punch was used where the corners of this $3 \times 3 \times 3$ pixel cube were excluded. In addition, far regions of reciprocal space, $\sqrt{h^2+k^2+l^2} > 14$, in which very little or no data had been measured was removed. The holes left by the Bragg peak punching process were filled with a smoothing function using the astropy convolution function for Python [6,7], where a Gaussian kernel with a standard deviation of two pixels was used. With this, the magnetic diffuse scattering has been isolated and the corresponding 3D-mΔPDF can be calculated as its Fourier transform. In addition to reconstructing the diffuse scattering within the punched regions, the convolution of the entire pattern with a Gaussian means that the Fourier transform yields the product of the 3D-mΔPDF and the Fourier transform of the kernel. Thus, the final 3D-mΔPDF is obtained by dividing the Fourier transform of the smoothed, isolated magnetic diffuse scattering with the Fourier transform of the Gaussian kernel. The process of isolating the magnetic diffuse scattering is shown in Figure S1.

### S2.2 3D-mΔPDF integration

Spherical integration boxes were used to get the intensities of the individual correlation peaks. The 3D-mΔPDF vas inspected to determine which range of sphere radii were appropriate to test for the data. The upper bound was determined to be 7 pixels as this is the maximal radius without overlap between adjacent integration spheres, while a lower bound of 5 pixels was chosen to avoid excluding real peak intensity. Fig. S2 shows an overlay of the different size integration spheres on the z=0 Å plane of the 1.5 K 3D-mΔPDF. From Fig S2(c) it is clear that a pixel radius of 7 includes significant noise in the integration and was thus excluded from consideration. Using a r=5 and 6 pixels resulted the same qualitative picture but a r=6 pixels yielded a slightly more believable temperature behavior and was thus chosen as the integration sphere radius. Fig. 4(f)-(g) from the main text is reproduced using a 5 pixel integration sphere radius in Fig. S3 showing the same qualitative, and similar quantitative, picture but a discontinuity like anomaly is observed the temperature dependence of nn14 around 17 K (Fig. S3(b)) resulting in the choice of a integration sphere pixel radius of 6 instead of 5.

## S3. Heat capacity modeling

The following expression was used to model the heat capacity as a function of temperature

$$C_p(T) = D(T) + 3E_1(T) + 3E_2(T) + S(T), \tag{S1}$$

where the first term is a Debye term, the second and third Einstein terms and the last a two-level Schottky term. The Debye term was defined as

$$D(T) = 9R\left(\frac{T}{\Theta_D}\right)^3 \int_0^{\frac{\Theta_D}{T}} \frac{x^4 e^x}{(e^x - 1)^2} dx, \tag{S2}$$

where R is the gas constant and $\Theta_D$ the Debye temperature. The Einstein term was defined as

$$E(T) = 3R\left(\frac{\Theta_E}{T}\right)^2 \frac{e^{\frac{\Theta_E}{T}}}{\left(e^{\frac{\Theta_E}{T}} - 1\right)^2}, \tag{S3}$$

where $\Theta_E$ is the Einstein temperature. The Schottky term was defined as

$$S(T) = R\frac{g_0}{g_1}\left(\frac{\delta_S}{T}\right)^2 \frac{1}{\left(1 + \frac{g_0}{g_1}e^{\frac{\delta_S}{T}}\right)^2}, \tag{S4}$$

where $g_0$ is the degeneracy of the ground state, $g_1$ the degeneracy of the excited state, and $\delta_S$ the energy barrier between the two states. These terms are actually defined for the heat capacity at constant volume, however, at the low temperatures used here the difference between $C_p$ and $C_v$ are negligible. This difference can, to first order, be approximated by the addition of a linear and a quadratic term in $T$ [8] Inclusion of these in the description of the heat capacity did not improve the description of the lattice contribution, nor affect the other refined parameters, confirming that in the given temperature range the approximation $C_p \approx C_v$ is reasonable.

The heat capacity $C_p$ in the interval 2-300 K (Fig. S4) lacks any sharp peak with a discontinuity in $dC_p/dT$, as would be expected for a paramagnetic to antiferromagnetic transition, but does show a distinct anomalous feature at low temperature in the form of a broad peak with a maximum just above 10 K. The lattice contribution can reasonably be described with the sum of one Debye and two Einstein terms, similar to King [9] and Westrum & Grimes [10], however, below 50 K this description breaks down due to the presence of the anomaly. The $C_p$ curve resembles the heat capacity data reported by Friedberg [11], where a similar broad low-temperature peak was observed, matching closely both in position, shape, and magnitude. That peak bore a resemblance to a Schottky anomaly, which is characteristic feature of a system with different states separated by energy-barriers on the order of $k_B T$ [12]. However, since $ZnFe_2O_4$ at the time was established as an antiferromagnetic compound with a Néel temperature that matched the peak position, it was attributed to a paramagnetic to antiferromagnetic transition. The broadness of the feature was attributed to three possible factors: non-uniform composition, partial inversion, and a non-stoichiometric sample. Using energy dispersive X-ray spectroscopy, micro X-ray fluorescence spectroscopy and single crystal neutron diffraction data measured on the TOPAZ instrument at SNS [4] we show that the present $ZnFe_2O_4$ crystals are stoichiometric, homogenous and exhibit a low degree of inversion, dispelling these factors as the cause of the anomaly (see below for further details). Inclusion of a Schottky term in the model of the $C_p$ yields a reasonable description of the anomaly considering the simplicity of the model. The Debye and Einstein temperatures are in reasonable agreement with literature values [9,10].

A linear fit of the region $T < 5$ K is included in the inset to highlight the linear low temperature dependence of $C_p$. This is not expected for a Schottky anomaly but it is a signature of a spin-glass where the transition itself is not associated with any distinct feature [13,14].

## S4. Crystal quality

To determine the stoichiometry and possible compositional gradients in the crystals, a roughly 2 mm crystal was cut in half allowing for measurements through the entire crystal cross-section. Sample stoichiometry was examined using a micro X-ray fluorescence (micro-XRF) spectrometer, where 12 areas of $200 \times 200$ μm² distributed in three randomly selected zones of the cross-section of the crystal were measured. This yielded an average molar ratio of $Fe_2O_3$ to ZnO of $n_{Fe_2O_3} \cdot n_{ZnO}^{-1} = 1.06(2)$ and the resulting chemical formula $Zn_{0.96(1)}Fe_{2.04(1)}O_4$.

Energy dispersive X-ray spectroscopy (EDX) images measured using a scanning electron microscope (SEM) on a ZnFe$_2$O$_4$ single crystal cross-section shows an even distribution of Zn and Fe confirming the lack of compositional gradients in the crystals (Fig. S5). The crystal was cast in epoxy for the process of cutting it in half and it is small pieces of this epoxy on the surface causing the carbon signal.

Structural refinement of the TOPAZ data with the total stoichiometry fixed to the micro-XRF results revealed a degree of inversion of 3.0(2) %, see Table S1 for details.

**Table S1.** Crystallographic data and collection parameters for ZnFe$_2$O$_4$ measured at TOPAZ

|  | ZnFe$_2$O$_4$ |
|---|---|
| Temperature (K) | 90 |
| Crystal system | Cubic |
| Space group | $Fd\bar{3}m$, no. 227 |
| $a$ (Å) | 8.4421(2) |
| $V$ (Å$^3$) | 601.65(3) |
| $Z$ | 8 |
| Radiation | Neutrons, time-of-flight Laue |
| Radiation wavelength (Å) | $0.40 - 3.47$ |
| Min, max transmission | $0.8925, 0.9225$ |
| $\mu(\lambda)$ (cm$^{-1}$) | $0.5905 + 0.0462\lambda$ |
| $d_{\min}$ (Å) | 0.357 |
| Total no. of reflections | 4871 |
| Rejected outliers ($|F_{obs} - F_{calc}| > 10\sigma(F_{obs})$) | 72 |
| $R_1, F^2$ | 0.0383 |
| $wR_2, F^2$ | 0.1021 |
| GooF | 2.67 |

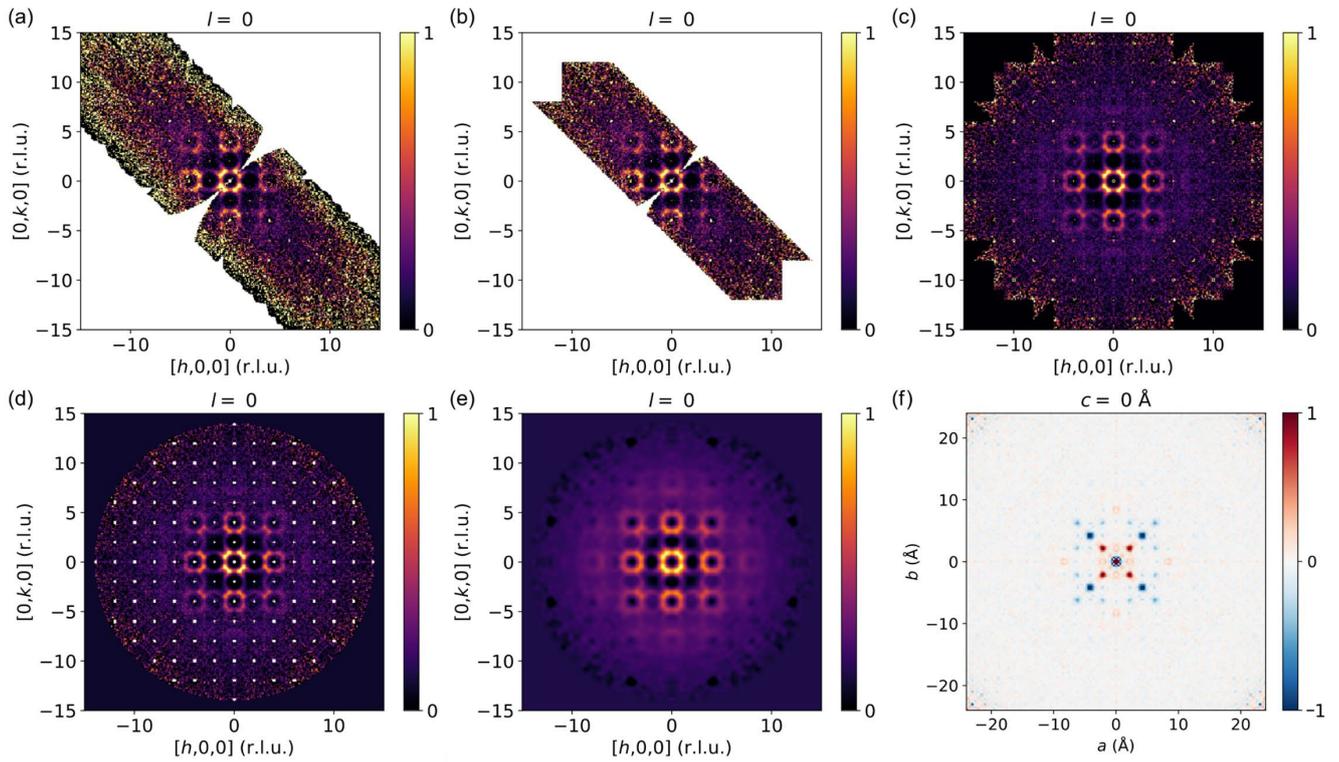

FIG. S6. (a) A slice of the 1.5 K scattering data after subtraction of 270 K dataset. (b) the same slice after removal of noisy edges. (c) The pattern after symmatrization according to the symmetry operations of the Laue class ($m\bar{3}m$). (d) The pattern after punching of Bragg peaks and elimination of regions where $\sqrt{h^2 + k^2 + l^2} > 14$. (e) The pattern after convolution with a Gaussian kernel. (f) The 3D-mΔPDF as the Fourier transform of e divided by the Fourier transform of the Gaussian kernel.

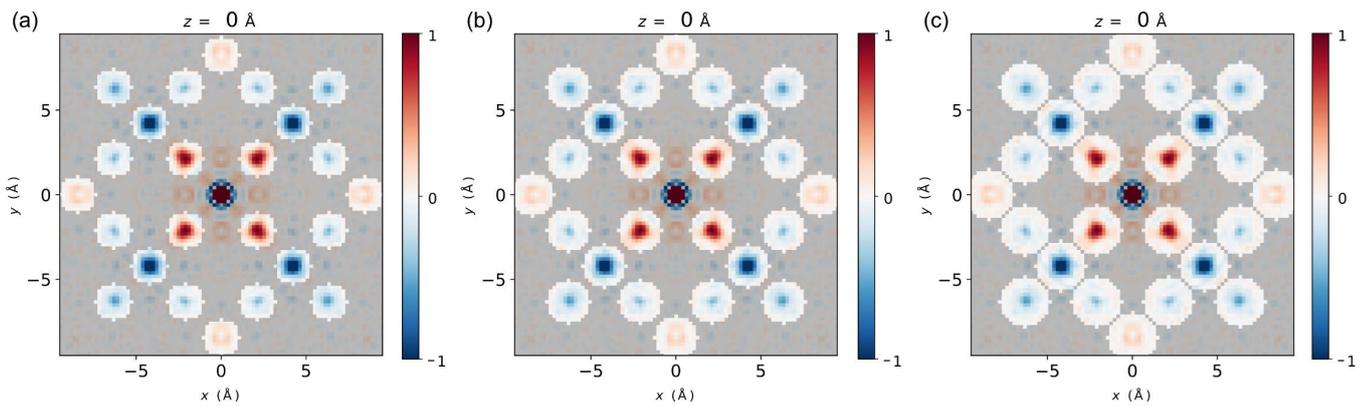

FIG. S7. The different integration sphere sizes overlayed on the *xy*-plane at *z*=0 Å of the 1.5 K 3D-mΔPDF. (a) r=5 pixels, (b) r=6 pixels, and (c) r=7 pixels.

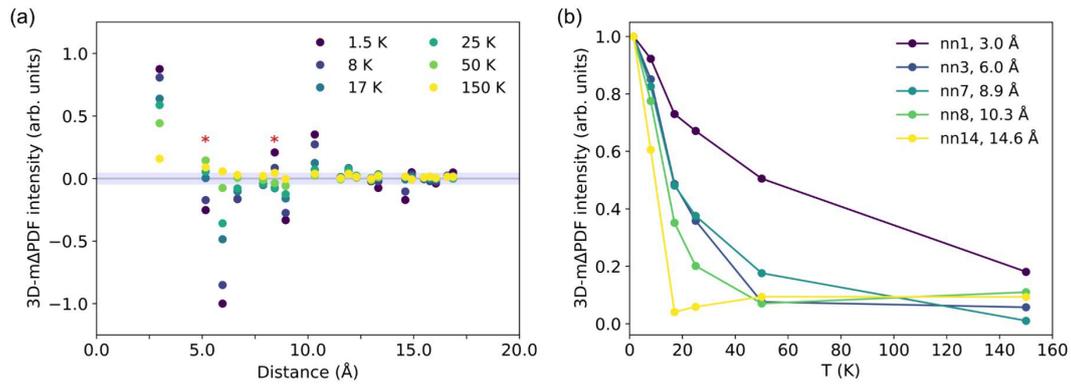

FIG. S8. Same as Fig. 4(f)-(g) from the main text but redone with a 5 pixel integration sphere radius. The qualitative agreement between these are clear, however, a peculiar feature resembling a discontinuity is seen in (b) for nn14 at 17 K which is not observed in Fig. 4(g) for which reason a pixel radius of 6 was chosen for the integration of the 3D-mΔPDF.

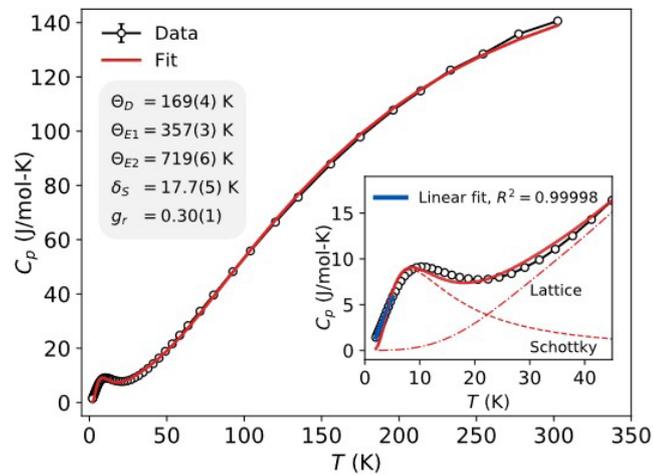

FIG. S4. The heat capacity as a function of temperature fitted using Eq. S1. The inset shows a zoom of the low temperature region with the broad anomaly where a linear fit of the sub 5 K data highlights the linear low temperature thermal behavior.

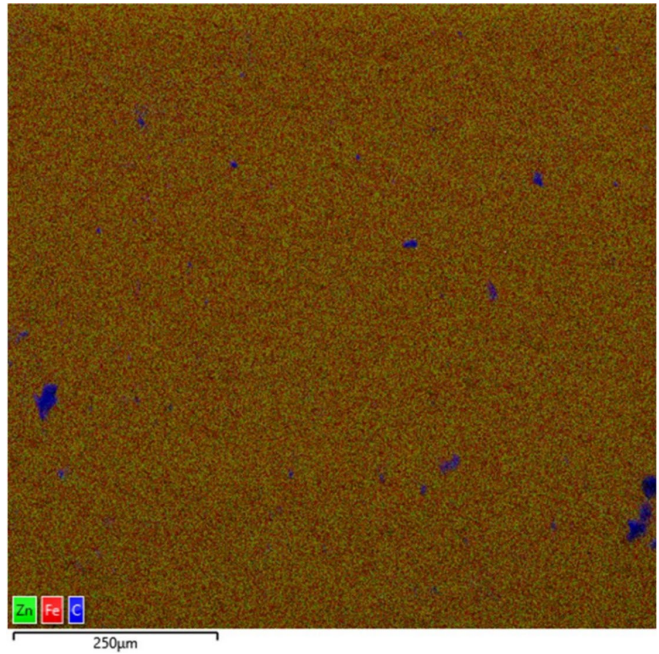

FIG. S5. SEM-EDX image of ZnFe$_2$O$_4$ single crystal cross section showing a homogenous distribution of Zn and Fe in the material. The carbon signal originates from epoxy used in the crystal cutting process.